\documentclass[12pt]{iopart}

\usepackage{epsfig}

\begin{document}

\title[Effective restoration of chiral and axial  symmetries]{Effective restoration of chiral and axial  symmetries at finite  temperature and density}

\author{M C Ruivo$^1$, P Costa$^1$,  C A de Sousa$^1$ and Yu L Kalinovsky$^{2,3}$}
\address{$^1$\ Departamento de F\'{\i}sica, Universidade de Coimbra,
P-3004-516 Coimbra, Portugal}
\address{$^2$\ Universit\'{e} de Li\`{e}ge, D\'{e}partment de Physique B5, Sart Tilman, B-4000, LIEGE 1, Belgium}
\address{$^3$\ Laboratory of Information Technologies,
Joint Institute for Nuclear Research, Dubna, Russia}

\ead{maria@teor.fis.uc.pt, pcosta@teor.fis.uc.pt,  celia@teor.fis.uc.pt and kalinov@qcd.phys.ulg.ac.be}

\begin{abstract}

The effective restoration of chiral and axial symmetries is investigated 
within the framework of the $SU(3)$ Nambu-Jona-Lasinio model. The 
topological susceptibility, modeled from lattice data at finite
temperature, is used to extract the  temperature dependence of the
coupling strength of the anomaly. The study of the scalar and pseudoscalar
mixing angles is performed in order to discuss the evolution of the flavor
combinations of $q \bar q$  pairs and its consequences for the
degeneracy of chiral partners. A similar  study at zero temperature and
finite density is also realized.
\end{abstract}
Understanding  properties of   low-lying hadron spectrum from the viewpoint of QCD dynamics and symmetries has been a  challenging question in  physics of strong interactions. In this context, the explicit and spontaneous breaking of chiral symmetry, as well as  the U$_A$(1) anomaly, play a special role, allowing for the conventional  assumption of low energy QCD: the octet of the low-lying pseudoscalar mesons ($\pi,\,K,\,\eta)$ consists of  approximate Goldstone bosons. 

It is generally expected that ultra-relativistic heavy-ion experiments will provide the strong interaction conditions which will lead to new physics. Restoration of symmetries and deconfinement  are expected to occur under those conditions, allowing for the search  of signatures of quark gluon plasma.     
It is also believed that at high temperatures the instanton effects are suppressed  due to the Debye-type screening \cite{Gross}. Then an effective restoration of U$_A$(1) symmetry is expected to occur at high temperatures.
In this context, it has been argued that the mass of the $\eta^\prime$ excitation in  hot and dense matter should be small, being expected the return of this "prodigal Goldstone boson" \cite{Kapusta}.

There should be other indications of the restoration of the axial symmetry,
like the vanishing of the topological susceptibility, which, in pure color
$SU(3)$ theory, can be linked to the $\eta'$ mass through the
Witten-Veneziano formula \cite{Veneziano}. In addition, since the presence of the axial
anomaly causes flavor mixing, with the consequent violation of the
Okubo-Zweig-Iizuka (OZI) rule, both for scalar and pseudoscalar mesons,
restoration  of axial symmetry should have relevant consequences for the
phenomenology of meson mixing angles, leading to the recovering of ideal
mixing.

We perform our calculations in the framework of a  SU(3) Nambu--Jona-Lasinio model  with a  Lagrangian density that includes the 't Hooft  interaction term:
\begin{eqnarray}
{\mathcal L\,}&=& \bar q\,(\,i\, {\gamma}^{\mu}\,\partial_\mu\,-\,\hat m)\,q
+ \frac{1}{2}\,g_S\,\,\sum_{a=0}^8\, [\,{(\,\bar q\,\lambda^a\, q\,)}
^2\,\,+\,\,{(\,\bar q \,i\,\gamma_5\,\lambda^a\, q\,)}^2\,] \nonumber\\
&+& g_D\,\{\mbox{det}\,[\bar q\,(1+\gamma_5)\,q] +\mbox{det}
\,[\bar q\,(1-\gamma_5)\,q]\}. \label{1}
\end{eqnarray}
By using a standard hadronization procedure,  an effective meson action is obtained, leading to gap equations for the constituent quark masses and to meson propagators from which several observables are calculated  \cite{costa}.
\vskip0.2cm
In what follows we will concentrate first on the restoration of the symmetries at zero density and finite temperature, and   the manifestations of the restoration on the scalar and pseudoscalar meson observables will be
analyzed. 
In chiral models, when the coefficient of the anomaly term in the Lagrangian ($g_D$ in the present case) is constant, in spite of the decreasing of the  U$_A$(1) violating quantities, this symmetry is not restored due to the fact that the strange quark condensate does not decrease enough. However, an effective restoration may be achieved by assuming that the strength of the anomaly coefficient is  a dropping function of the temperature \cite{kuni,Bielich,Ohta}.
The analysis  of the temperature dependence of the mixing angles, allowing for the understanding of the evolution of meson quark content and, in particular,  of the role of the strange order parameter in SU(3) chiral partners like $(\sigma,\eta)$ and $(f_0,\eta^\prime)$, can provide further indication of the restoration of the axial symmetry.    
\begin{figure}[h]
\includegraphics[width=18pc]{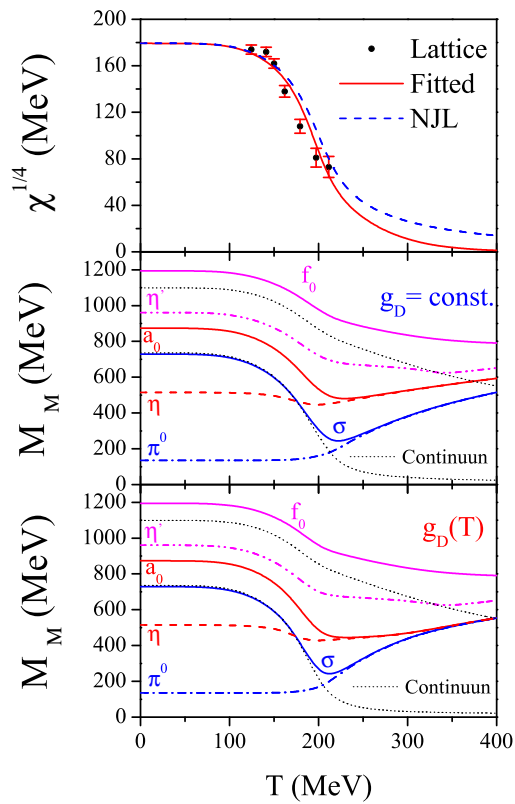}\hspace{-5pc}%
\begin{minipage}[b]{18pc}\caption{\label{label}Topological susceptibility (upper panel) from lattice data plotted with error bars \cite{lattice}. The solid (dashed) line represents our fitting with constant (temperature dependent) $g_D$. Temperature dependence of the meson masses, as well as that of the continuum thresholds $2M_u$ and $2M_s$ with constant (middle  panel) and temperature dependent (lower panel) $g_D$.}
\end{minipage}
\end{figure}
In the present work, following the methodology of \cite{Ohta}, we  extract the temperature dependence of   the anomaly coefficient  $g_D$  from the lattice results for the topological susceptibility  \cite{lattice} (see Fig. 1, upper panel).


Our results, concerning the restoration of the chiral phase transition, at zero density and finite temperature, indicate, as usual, a smooth crossover: at temperatures around $T\approx200$ MeV the mass of the light quarks drops to the current quark mass. The strange quark mass also starts to decrease significantly in this temperature range, however even at $T = 400$ MeV it is still 2 times the strange current quark mass. 
In fact, as $m_u=m_d<m_s$, the (sub)group SU(2)$\otimes$SU(2) is a much better symmetry of the Lagrangian (1). 
So, the effective restoration of the above symmetry  implies the
degeneracy between the  chiral partners $(\pi,\sigma)$ and $(a_0,\eta)$ around $T\approx250$ MeV (see Fig. 1, lower panel).
For temperatures  about $T\approx350$ MeV, both $a_0$ and $\sigma$ mesons become degenerate with the $\pi$ and $\eta$ mesons, showing an effective restoration of both chiral and axial symmetries.
In fact, the U$_A$(1) symmetry is effectively restored when the U$_A$(1) violating quantities show a tendency to vanish, which means that the four meson masses are degenerated and the topological susceptibility goes to zero. Without the restoration of U$_A$(1) symmetry, the $a_0$ mass was moved upwards and never met the $\pi$ mass as can be seen in Fig. 1, middle panel. The same argument is valid for $\sigma$ meson comparatively with the $\eta$ meson mass. We remember that the determinant term acts in an opposite way for the scalar and pseudoscalar mesons.
So, only after the effective restoration of U$_A$(1) symmetry we can recover the SU(3) chiral partners $(\pi,a_0)$ and $(\eta,\sigma)$ which are now all degenerated.
However, the $\eta^\prime$ and $f_0$ masses do not yet show a clear tendency to converge in the region of temperatures studied.

In order to understand this behavior, we   also analyze the temperature dependence of the mixing angle: $\theta_S$ starts at $16^{\circ}$ and goes, smoothly, to the ideal mixing angle $35.264^{\circ}$ and $\theta_P$ starts at $-5.8^{\circ}$ and goes to the ideal mixing angle $-54.7^{\circ}$. 
This means that flavor mixing no more exists.
In fact, analyzing the behavior of the SU(2) chiral partner ($\eta,a_0$) with the temperature (Fig. 1, lower panel), we found that the $a_0$ meson is always a purely non strange $q \bar q$  system while the $\eta$ meson, at $T = 0$ MeV, has a strange component and becomes purely non strange when $\theta_P$ goes to $-54.7^{\circ}$ at $T \approx 250 $ MeV. At this temperature they start to be degenerated. Concerning the  SU(2) chiral partner ($\pi^0,\sigma$), at $T = 0$ MeV, $\pi^0$ is always a light quark system and the $\sigma$ meson has a strange component,  but becomes purely non strange when $\theta_S$ goes to $35.264^{\circ}$ at $T \approx 250 $ MeV. In summary,  we see that the U$_A$(1) symmetry is effectively restored at $T\approx350$ MeV:   $\pi^0$, $\sigma$, $\eta$ and $a_0$ mesons become degenerated, the OZI rule is verified and $\chi$ goes asymptotically to zero.
We remember that  $\pi$, $\sigma$, $\eta_{\rm ns}$ and $a_0$ form a complete representation of U(2)$\otimes$U(2) symmetry, but the $\pi$ and the $a_0$ do not belong to the same SU(2)$\otimes$SU(2) multiplet. 
The partners $(f_0\,,\eta')$, which become purely strange at high
temperatures,  do not converge probably due to the fact that chiral symmetry is not restored in the strange sector.
\vskip0.2cm

Due to recent studies on lattice QCD at finite chemical potential,  it is tempting to investigate also the restoration of the U$_A$(1) symmetry at finite  density and zero temperature. The disadvantage in this case is that there are no firmly lattice results for the density dependence of the topological susceptibility, to be used as input, although preliminary results indicate that there is a drop of $\chi$ with increasing baryonic matter \cite{latticed}.
 We postulate a dependence for the topological susceptibility formally similar to the temperature case and assume  that the coefficient of the anomaly is a dropping function of the baryonic density.
Here  we   consider quark matter simulating "neutron" matter. This "neutron" matter is in chemical equilibrium, maintained by weak interactions and with charge neutrality, and undergoes a first  order phase transition  \cite{costa,costabig}.
As in the temperature case, the non strange quark mass decreases sharply, reflecting the approximate restoration of  SU(2) chiral symmetry, and there are no signs of restoration of symmetry in the strange sector.
Differently to the finite temperature case, where the mixing angles have a monotonous behavior, we found that at finite density the pseudoscalar mixing angle is very sensitive to the fraction of strange quarks present in the medium, eventually changing sign when this fraction is small. It is what happens in the present case where the 
sign changes at  $\rho_B\approx 4\,\rho_0$. This leads  to a change of identity between $\eta$ and $\eta'$ and, consequently,  to the degeneracy of   $\pi^0$ with $\eta^\prime$.
Although this difference, the overall conclusions are  qualitatively similar in both cases \cite{costanew}. 
The main difference comes from the description of mesons which,  differently from the non zero temperature case,  are all bound states at high densities.

In summary, the present investigation explores a framework to study effective chiral and axial symmetry restoration  with temperature and density. 
We implement  a framework based on  a lattice-inspired behavior of the topological susceptibility to extract  the coupling strength  of the anomaly.  The evolution of the flavor combination of $q\bar q$ pairs, and the consequent convergence of appropriate chiral partners provide the criterion to identify an effective restoration and chiral and axial symmetries. 
Talking about the   U(3)$\otimes$U(3) symmetry, we do not observe the full restoration of this symmetry, although its axial part is restored at a moderate temperature (density).
However, the role of U$_A$(1) symmetry for finite temperature, and mainly for finite density media, has not been so far investigated and this question is still controversial and not settled yet.
We hope that new studies, especially lattice based and experimental ones, can finally clarify it. 

\vskip0.2cm
Work supported by grant SFRH/BD/3296/2000 (P. Costa), by grant RFBR 03-01-00657, Centro de F\'{\i}sica Te\'orica and GTAE, and by FEDER/FCT under projects POCTI/FIS/451/94 and POCTI/FNU/50326/2003.

\end{document}